# Shape-Dependent Oriented Trapping and Scaffolding of Plasmonic Nanoparticles by Topological Defects for Self-Assembly of Colloidal Dimers in Liquid Crystals

Bohdan Senyuk,[†] Julian S. Evans,[†,‡] Paul J. Ackerman,[†] Taewoo Lee,[†] Pramit Manna,[§] Leonid Vigderman,[§] Eugene R. Zubarev,[§] Jao van de Lagemaat,[‡,∥] and Ivan I. Smalyukh[*,†,∥]

[†]Department of Physics, Materials Science and Engineering Program, Department of Electrical, Computer, and Energy Engineering, and Liquid Crystals Materials Research Center, University of Colorado at Boulder, Boulder, Colorado 80309, United States

[‡]National Renewable Energy Laboratory, Golden, Colorado 80401, United States

[§]Department of Chemistry, Rice University, Houston, Texas 77005, United States

[∥]Renewable and Sustainable Energy Institute, National Renewable Energy Laboratory and University of Colorado at Boulder, Boulder, Colorado 80309, United States

*E-mail: ivan.smalyukh@colorado.edu.

**Abstract**

We demonstrate scaffolding of plasmonic nanoparticles by topological defects induced by colloidal microspheres to match their surface boundary conditions with a uniform far-field alignment in a liquid crystal host. Displacing energetically costly liquid crystal regions of reduced order, anisotropic nanoparticles with concave or convex shapes not only stably localize in defects but also self-orient with respect to the microsphere surface. Using laser tweezers, we manipulate the ensuing nanoparticle-microsphere colloidal dimers, probing the strength of elastic binding and demonstrating self-assembly of hierarchical colloidal superstructures such as chains and arrays.

---

Starting from the development of facile synthetic methods, gold nanoparticles have been the focus in the quest for understanding and exploiting their extraordinary optical properties arising from the localized surface plasmon resonance (SPR).[1–3] Strong absorption and local field enhancement observed in such nanoparticles are of great interest for applications in nanophotonics, solar cells,[4] photothermal therapies, and imaging.[2,3] Nonspherical particles can exhibit SPR tunable throughout the visible and near-infrared spectral regions[3,5] and their plasmonic properties can be further enhanced by the collective behavior when the interparticle separation becomes comparable to their size.[2] For example, multipole SPRs[6] and well-defined assemblies[7] of nanoparticles give simple ways of obtaining a nontrivial magnetic response, making plasmonic colloidal systems useful for the preparation of metamaterials.[8] Near-field proximity of plasmonic nanoparticles can also alter the fluorescent behavior of semiconductor and dielectric particles as well as molecular dyes.[9–11] Precise arrangement of metal nanoparticles can yield highly desirable functionality for applications such as nanoantennas.[12–14] These examples demonstrate the great potential of noncontact manipulation and trapping techniques that would allow for precise positioning of plasmonic and other particles with respect to each other in fluid host media.

Although spatial manipulation of colloids of most compositions can be accomplished using optical trapping, this technique requires continuous focusing of high-intensity laser beams and has diffraction-limited minimum interparticle distance (typically hundreds of nanometers) at which two particles can be stably localized with respect to each other. Furthermore, optical trapping of metal nanoparticles[15–18] is limited by heating,[17] the relationship between trapping and SPR wavelengths,[18] poor control of anisotropic particle orientation, and other factors. In anisotropic fluids such as liquid crystals (LCs),[19] which are of great interest for introducing tunability into the dispersions of plasmonic nanoparticles,[12,20–22] manipulation becomes even more complicated as the electric field of the trapping beam can cause local realignment and induce phase transitions in the LC host.

In this Letter, we use topological singularities (defects) for elastic scaffolding of small plasmonic nanoparticles next to bigger colloidal microspheres. These microparticle-induced defects in a nematic LC elastically trap nanoparticles with concave and convex anisotropic shapes. Nanoparticles are attracted to defects to displace the energetically costly defect core regions of reduced order parameter surrounded by strong long-range elastic distortions of the LC. By means of dark-field microscopy and polarization-sensitive two-photon luminescence (TPL) imaging, we demonstrate that anisotropic gold colloids such as nanoplatelets and elongated rodlike octagonal, pentagonal, and starfruit-like prisms not only self-localize within topological defects but also exhibit well-defined orientation with respect to the neighboring microparticle and the uniform far-field LC alignment. Using laser tweezers, we manipulate the dielectric colloidal microspheres along with the defect-bound plasmonic nanoparticles next to them. We explore the strength of nanoparticle-defect interaction forces and show hierarchical self-assembly of metal and dielectric colloids into chains and two-dimensional arrays. The demonstrated control of position and orientation of the plasmonic nanoparticles allows for the creation of new nanoscale systems with interesting potential applications in the areas of nanoscale imaging and nanoscale energy conversion.

We use a single-compound LC material 4-pentyl-4′-cyanobiphenyl (5CB, obtained from Frinton Laboratories, Inc.) and nanoparticles with different anisotropic shapes, dimensions (characterized by the axial diameter $D_{np}$ and length $L_{np}$), and surface chemistry (Table 1). Ethanol dispersions of microgolds

(MGs), starfruits (SFs), nanobursts (NBs) and nanorods (GNRs) with NSol(akyl acrylate) polymer conjugation (Figure 1a−d) were obtained from Nanopartz, Inc. (Colorado, U.S.A.). The smallest gold nanorods (sGNRs) with polystyrene capping were obtained by means of an esterification reaction between mercaptophenol-functionalized nanorods and carboxyl-terminated polystyrene (MW = 5000 g/mol) as described elsewhere.[23,24] Silica microspheres of 3 μm in diameter were obtained from Duke Scientific as a powder and treated with a surfactant [3-(trimethoxysilyl)propyl]octadecyl-dimethylammonium chloride (DMOAP) to set vertical surface boundary conditions for the LC molecular alignment.[25] Melamine resin spherical particles of 2 μm in diameter and with tangential surface boundary conditions[26] were obtained in an aqueous dispersion (from Sigma-Aldrich) and turned into a powder via slow water drying. These microparticles were then dispersed in the LC by direct mixing and a 15−30 min sonication to break apart pre-existing colloidal aggregates. Gold nanoparticles were dispersed from ethanol to toluene and then added to the mixture of LC and microspheres; toluene was evaporated afterward. This colloidal dispersion was sonicated for 30−60 min while slowly changing the samples' temperature from isotropic to the LC state to obtain isolated microspheres and metal nanoparticles in the nematic host. Using capillary forces, LC colloidal dispersions were filled into ∼5−10 μm thick cells formed by substrates spaced by glass fiber spacers and treated to provide planar or vertical boundary conditions. In order to set homogeneous planar boundary conditions, glass plates were spin-coated with polyimide PI2555 (HD Micro-System) on their inner surfaces, unidirectionally rubbed, and then glued together by epoxy while aligned to have "antiparallel" rubbing directions at the opposite plates. Surface treatment of substrates with DMOAP was used to set the vertical surface boundary conditions for the LC molecules.

We have used dark-field, bright-field, and polarizing modes of optical imaging with an inverted Olympus microscope IX81 integrated with a holographic optical trapping (HOT) setup[27] operating at wavelength λ = 1064 nm. The strong scattering of metal nanoparticles[28] in the dark-field microscopy[20,29] (Figure 1e−g) allowed us to visualize individual colloids of size smaller than the diffraction limit. The adjustable numerical aperture (NA) of the high magnification oil objective 100× was selected to be smaller (NA = 0.6) than the NA of the dark-field condenser ($NA_{cond}$ = 1.2). The interaction between gold nanoparticles and topological defects was monitored with a PointGrey CCD camera at a frame rate of 15 fps using dark-field microscopy, enabling a direct measurement of elastic interaction forces.[29]

Orientation-sensitive imaging of sGNRs and GNRs was performed using their polarized TPL.[30,31] We used a tunable (680−1080 nm) Ti:sapphire oscillator (140 fs, 80 MHz, Chameleon Ultra-II, Coherent) to generate TPL from sGNRs and GNRs (Figure 1h,i). To enhance the polarization dependence of TPL, the excitation of gold nanorods was performed at wavelengths in the proximity of the longitudinal SPR peak (850 nm for sGNR and ∼650 nm for GNR), and the TPL signal from a single nanorod was detected within the spectral range of 400−700 nm (Supporting Information) by a photomultiplier tube (H5784-20, Hamamatsu). The in-plane position of the focused excitation beam was controlled by a galvano-mirror scanning unit (FV300, Olympus) and its polarization was varied using a half-wave retardation plate mounted immediately before the 100× oil objective. The average power of the fast-scanning excitation beam in the sample plane was ∼100−200 μW (see Supporting Information for details), low enough to be insufficient to change the temperature of LC and affect its local ordering due to the laser-induced heating of gold nanoparticles.[17]

The unique property of colloidal dispersions of all studied anisotropic gold nanoparticles in LCs is that they exhibit well defined alignment with respect to the uniform far-field director $\mathbf{n}_0$ that can be controlled on large scales via treatment of confining substrates (like in flat-panel LC displays). Nanoparticle shape and capping can promote vertical or tangential alignment of the LC molecules at its surface. Particles disturb the initially uniform average local molecular alignment along the LC director $\mathbf{n}$ describing the average local orientation of the rod-like molecules,[19] and induce spatial distortions described by the director field $\mathbf{n}(\mathbf{r})$ and dependent on the type and strength of anchoring at their surfaces. The ensuing equilibrium orientation of anisotropic nanoparticles depends on their size and the type and strength of surface anchoring and can be characterized by a dimensionless parameter[32,33] $\omega \sim R_{np}W_a/K$, where $W_a$ is a surface anchoring coefficient at the nanoparticle-nematic interface, $R_{np} = D_{np}/2$ is a radius of nanoparticle, and $K$ is an average Frank elastic constant.[19] Taking typical values of $K \approx 10$ pN and $W_a \sim 10^{-6}-10^{-4}$ J/m$^2$ and sizes of studied nanoparticles, we find that $\omega \sim 0.001-1$ varies within a wide range. Although NSol coatings at flat surfaces promote vertical boundary conditions while polystyrene coatings promote planar boundary conditions at flat LC-solid interfaces, the resulting spatial distortions of $\mathbf{n}(\mathbf{r})$ due to nanoparticles also depend on their shape (i.e., concave vs convex). Since the ensuing $\mathbf{n}(\mathbf{r})$ structures appear on submicrometer scales, their details cannot be resolved by polarizing optical microscopy. However, by means of dark-field microscopy, we directly observe nanoparticles as bright diffraction-limited scattering spots on a dark background of the nonscattering aligned LC (Figures 1e−g). Dark-field imaging reveals the orientation of nanoparticles: pentagonal MG nanorods (Figure 1a) align perpendicular to the far-field director $\mathbf{n}_0$ (Figure 1e) while SFs align along $\mathbf{n}_0$ (Figure 1f) and NBs align with their large-area flat faces normal to $\mathbf{n}_0$ (Figure 1g).

The orientation of small octagonal sGNR and GNR nanorods was determined using polarized TPL (Figures 1h,i and 2). The TPL intensity from these particles depends on the polarization direction of the excitation beam as $\cos^4\beta$, where $\beta$ is the angle between the long axis of the nanorod and the polarization of the excitation light (see Supporting Information for details).[30] Figure 2 shows a polar plot of TPL intensity versus $\beta$ for a nanorod aligned along $\mathbf{n}_0$. The TPL intensity from both sGNR and GNR particles is at minimum for the polarization of excitation light perpendicular to $\mathbf{n}_0$ (Figure 1h) and maximum for their parallel orientation (Figure 1i), allowing us to unambiguously establish that sGNRs and GNRs align parallel to $\mathbf{n}_0$ (Figure 1m,n). Figure 1j−n show the corresponding schematics of $\mathbf{n}(\mathbf{r})$ and defects induced in the LC by studied gold nanoparticles, as established based on the known boundary conditions at their surface and the comparative analysis of different types of optical images, such as the dark-field and polarizing microscopy images shown in Figure 1f for the case of SFs shown in Figure 1b. MG nanoparticles (Figure 1a) have the shape of a pentagonal prism and align in the nematic LC with their long dimension $L_{np}$ normal to $\mathbf{n}_0$ (Figure 1j). Because of the homeotropic surface anchoring, they are encircled by a disclination loop of strength $k = -1/2$ (Figure 1j), where $k$ is defined as a number of director turns by $2\pi$ when the defect core is circumnavigated once. MGs freely rotate around $\mathbf{n}_0$. NB nanoplatelets (Figure 1c) with multiple irregular sharp edges and vertical surface boundary conditions align with their flat sides normal to $\mathbf{n}_0$, inducing a $k = -1/2$ disclination loop around their perimeter (Figure 1l). The SFs have complex elongated shape with the concave base in the form of a star (Figure 1b). They align with their long axes along $\mathbf{n}_0$ and, as can be seen from polarizing microscopy textures (inset of Figure 1f), induce $\mathbf{n}(\mathbf{r})$ of quadrupolar symmetry with two surface point defects called "boojums" at their ends (Figure 1k). This $\mathbf{n}(\mathbf{r})$ structure can be explained by the grooved surface relief of the rodlike SF nanoparticles with star-shaped base (Figure 1b). Bulk elastic energy is minimized when the LC director locally follows these grooves instead of satisfying antagonistic surface boundary conditions exerted by the complex ribbed surface relief of these nanoparticles (Figure 1f,k), explaining the observed $\mathbf{n}(\mathbf{r})$-configuration. GNR and sGNR particles have octagonal cross sections. GNRs promote vertical alignment of LC molecules at their surface and align with their long axes along $\mathbf{n}_0$ with the $k = -1/2$ disclination loop encircling them in the plane normal to their long axis (Figure 1m). Since $\omega$ is small, one can expect that this disclination loop is "virtual" or located at the particle surface.[32−34] The anchoring at sGNR polystyrene-capped surfaces is tangential, causing weak $\mathbf{n}(\mathbf{r})$-distortions of quadrupolar symmetry with two boojums (Figure 1n). These findings for complex-shaped colloids are consistent with the studies of colloidal cylinders in LCs,[32,33,35−37] which showed that their orientation can be both along and perpendicular to $\mathbf{n}_0$ for normal boundary conditions and that rods with tangential boundary conditions typically align parallel to $\mathbf{n}_0$.

To enable and control spatial localization of fluid-borne nanoparticles of varying sizes and shapes, we use colloidal microspheres also dispersed in the LC. These solid microspheres induce much stronger distortions of $\mathbf{n}(\mathbf{r})$ compared to the nanoparticles, causing well-defined singular point and line defects.[38,39] Depending on the strength,[40] the vertical surface anchoring at silica spheres causes either dipolar distortions of $\mathbf{n}(\mathbf{r})$ with a hyperbolic point defect[19] of topological charge $N = -1$ (Figure 3a,b,d,f) or quadrupolar distortions of $\mathbf{n}(\mathbf{r})$ with $k = -1/2$ disclination loop ("Saturn ring")[41] encircling the sphere along its equator in the plane normal to $\mathbf{n}_0$ (Figure 3i,j). In our experiments, spherical particles with the hedgehog defect were commonly observed in thicker (10 μm) thick cells and particles with the Saturn ring defect were observed predominantly in thinner (5 μm) thick cells.[42] In dark-field microscopy, the hyperbolic point defect is seen as a weakly scattering spot near a strongly scattering silica sphere (Figure 3a and Figure 4a, marked by a yellow arrow). The Saturn ring (Figure 3j) appears as a weakly scattering line that is clearly seen at the microsphere edges (Figure 3i, marked by a green arrow).

Colloidal particles suspended in LCs experience anisotropic interaction forces[38,39] that arise due to minimization of the elastic free energy driven by sharing of director distortions introduced by neighboring particles. Attraction of nanoparticles into topological singularities induced by microspheres is of similar nature and arises due to the displacement of energetically costly regions of distorted **n(r)** and "melted" isotropic defect cores by the nanoparticles. We have used the HOT system[27] to translate the microspheres to the vicinity of metal nanoparticles. Dark-field video microscopy then tracked the motion of colloids after releasing the microparticles in the vicinity of nanoparticles. The sGNRs, GNRs (Figure 3a,i), and NBs (Figure 4a) were attracted toward point and line defects within a couple micrometers and eventually trapped by them (Figures 3a,c,e,g,h and 4a−c). Figures 3a and 4a show such nanoparticles being attracted into the point defect in the direction roughly along **n**$_0$. The entrapment of nanorods by the Saturn ring defect around microsphere is demonstrated in Figure 3i,j. At large distances, the directionality of the elastic interactions between the microsphere and GNR is similar to that of two quadrupolar LC colloids with Saturn rings and the same curvature of **n(r)**-distortions,[38,39,42] repelling at center-to-center separation vectors parallel and perpendicular to **n**$_0$ but attracting at intermediate angles (Supporting Information Figure S4).[38,39,42] Even when released with center-to-center vector orientation corresponding to repulsion (Figure 3j), the GNR slowly drifts to the attractive angular sector and attracts toward the microsphere roughly along the expected direction of maximum attraction between such quadrupoles.[38,39,42] Once in the vicinity of the microsphere, the GNR slides around its surface by continuously displacing increasingly stronger director distortions and eventually localizes in the defect (Figure 3j). In contrast, the directionality of the elastic interactions between microsphere with the Saturn ring and sGNR with two boojums (Figure 3j) is similar to the case of two quadrupolar LC colloids having opposite distributions of defect signs and opposite curvature of **n(r)**-distortions,[43] attracting at center-to-center separation vectors parallel and perpendicular to **n**$_0$ but repelling at the intermediate angles (Supporting Information Figure S4). When released having the center-to-center separation vector perpendicular to **n**$_0$, sGNRs head straight to the disclination until entrapped (Figure 3j). Once trapped in the point defect, sGNRs and GNRs spontaneously orient orthogonally to the sphere surface and along **n**$_0$, as revealed by polarized TPL. GNRs and sGNRs trapped in disclinations align along the defect lines, minimizing the total free energy by maximizing the volume of melted defect core that they displace at this orientation.

Figure 4 shows that NB nanoplatelets are attracted to the point defect not only from close proximity (Figure 4a) but even when released at distant initial locations on the side of the microsphere opposite to that of the defect, moving down the path corresponding to the strongest gradient of $\mathbf{n}(\mathbf{r})$ (Figure 4e). Defect-trapped NBs orient to have their large-area faces locally parallel to the surface of the microsphere (Figure 4b,c and Supporting Information Figure S3). Several NBs can be collected in the region of the point defect simultaneously while preserving this orientation with respect to the microsphere (Figure 4e).

By tracking positions of colloids with video microscopy (Figure 4a), one can measure the time dependent center-to-center separation $r_{cc}(t)$ between the nanoparticle and the microsphere (inset of Figure 4d). In the regime of low Reynolds numbers (Re $\ll 10^{-7}$ in our case), the inertia effects are negligible and the attractive elastic force is balanced by the viscous drag force[37,44] $F_d = c_d v_r(t)$, where $c_d$ is an average drag coefficient in an anisotropic fluid and $v_r(t) = dr_{cc}/dt$ is a relative velocity of the nanoparticle with respect to the center of the microsphere. To estimate $c_d$, we have measured self-diffusion constants $D_\parallel$ and $D_\perp$ of the nanoparticle for the diffusion directions along and perpendicular to $\mathbf{n}_0$, respectively.[44] Using the estimated $c_d$ and measured $v_r(t)$ (see, for example, the inset in Figure 4d), one determines the separation-dependent attractive potential (Figure 4d), the minimum of which yields the attraction energy $W_0$ of a nanoparticle to the defect before being entrapped in the defect core; the interaction energies increase with the nanoparticle size (Table 1). MG and SF colloids also experience attraction toward the silica microsphere (Figure 5a). However, in contrast to GNRs, sGNRs, and NBs they do not self-localize into the hyperbolic hedgehog core but rather in the region of distorted $\mathbf{n}(\mathbf{r})$ at a distance to the point defect comparable to their size, forming weakly bound microsphere-nanoparticle pairs with the $r_{cc}$ changes due to thermal fluctuations within ~0.5 μm. The same is true in the vicinity of boojum defects induced by melamine resin microparticles (Figure 5e,f): SFs do not localize in the boojum regions, but stay elastically bound in the nearby region with strongly distorted $\mathbf{n}(\mathbf{r})$, with $r_{cc}$ also changing due to thermal fluctuations up to ~0.5 μm. Their behavior is consistent with the elastic dipole−quadrupole and quadrupole−quadrupole interactions that was previously explored for microspheres.[27,38,39,43] To locate MG and SF nanoparticles more precisely within the defects, we have used optical manipulation. Laser beams of moderate power (20−50 mW) focused between the point defect and the nanoparticle lead to trapping of MGs and SFs at the bulk point defects (Figures 5b,d) or surface boojums (Figures 5f,g). Dark-field microscopy (Figure 5b,f) reveals that defect-trapped MGs and SFs orient perpendicular to the microsphere surface.

The shape- and size-dependent oriented attraction and trapping of metallic nanoparticles by topological defects are driven by the minimization of director distortions and corresponding elastic energies. They can be qualitatively understood by comparing anisotropic particle dimensions $R_{np}$, $L_{np}$, and the so-called surface extrapolation length $l_a = K/W_a \sim 0.1{-}10$ μm.[19] The surface anchoring effects are negligible for particles with $R_{np} < l_a$, so that they are simply driven toward regions with the strongest gradients of $\mathbf{n}(\mathbf{r})$ and eventually trapped while displacing LC regions with increasingly stronger and stronger elastic distortions. One can roughly estimate the maximum free energy reduction $\Delta W \sim KR_{eqs}$ due to replacing the energetically costly volume of the distortions by a particle of volume $V_{np} = \pi D_{np}^2 L_{np}/4$ as the elastic energy of distortions in the region limited by an "equivalent" sphere of volume $V_{np}$ and radius $R_{eqs}$ (Table 1). As expected, the estimated energy reduction $\Delta W$ is larger or comparable to the attraction energy $W_0$ between the sGNR, GNR ($R_{eqs} < l_a$) and defects measured near the contact with the singularity at a distance limited by a ~10 nm resolution of videotracking (Table 1). When $R_{eqs} \gtrsim l_a$, the director field around the nanoparticle is highly dependent on the boundary conditions at its surface, and such a nanoparticle may induce additional elastic distortions if placed into the region with the nonuniform $\mathbf{n}(\mathbf{r})$, so that the superposition of pre-existing and nanoparticle-induced director distortions may not always lead to lower elastic energy. This is the case for MG and SF nanoparticles, which are driven toward the region with strongly distorted $\mathbf{n}(\mathbf{r})$ around the microparticles (Figure 5a) until the mismatch of $\mathbf{n}(\mathbf{r})$ around the microparticle and nanoparticles causes an energetic barrier $> k_BT$ and localizes them next to the topological singularity in a location that is likely a metastable state. Laser tweezers help to overcome this energetic barrier when the trapping beam is placed in the region between the point defect and MG or SF, further distorting $\mathbf{n}(\mathbf{r})$ and dragging MGs and SFs into the stable trapping position coinciding with the core of the point defect (Figures 5b,d,f,g). Although $R_{eqs}$ of NBs is comparable to that of MGs and SFs, NB colloids spontaneously localize in the defect cores, which is because the elastic distortions induced by these nanoparticles are weak and distributed around their perimeter, so that they do not result in significant elastic energy barriers as the particle traverses toward defect cores. This finding indicates that oriented trapping of nanoparticles by topological defects is highly dependent on their shape, in addition to the particle size.

Topological singularities entrap nanoparticles with nanoscale precision on the order of the defect core radius (~10 nm) with respect to the center of the defect core, better than in the conventional optical traps (dependent on laser power but on the order of the diffraction limit, i.e., about hundreds of nanometers) and even better than in the case of plasmonic optical traps.[45] The changes of the center-to-center distance between the microsphere and entrapped nanoparticles are smaller or comparable to the 10 nm resolution with which we can localize the centers of these colloids by means of video microscopy. At short distances $< R_{np}$, the topological defect trapping force $F_{tdt}$ is expected to increase linearly with the displacement[26,27] $\Delta r$ of the nanoparticle from the equilibrium position in the elastic trap following Hooke's law $F_{tdt} = \kappa_{tdt}\Delta r$, where $\kappa_{tdt}$ is a trap stiffness. $F_{tdt}$ is maximized when $\Delta r \sim R_{eqs}$, and can be called a "trap escape force" $F_{te}$ in analogy with the case of optical trapping.[26,27] This escape force for the point defect trap can be determined using the energy $\Delta W_R$ cost of displacing the nanoparticle from the elastic trap by a distance $R_{eqs}$. The reduction energy $\Delta W_R$ is equal to the sum of the defect core energy and the elastic energy of director distortions in a volume of the defect occupied by the nanoparticle. In the case of the hyperbolic point defect (Figure 3f), $\Delta W_R$ can be estimated using the expression $W_{hp} = 8\pi K(R_{eqs} - r_{pc})/3 + \gamma r_{pc}^3$, where $W_{hp}$ is the elastic energy of the hedgehog defect, $\gamma \sim 10^5$ J/m$^3$ is the energy density of the isotropic core and $r_{pc} = [(8/9)(\pi/\gamma)K]^{1/2} \approx 13$ nm is its radius.[19] The first term in $W_{hp}$ corresponds to the elastic contribution and the second term describes the energy of a "melted" isotropic core. In the case of the Saturn ring defect (Figure 3i,j), $\Delta W_R$ can be estimated as an elastic energy of a wedge disclination loop of the strength $k = -1/2$, $W_{sr} = \pi k^2 KL \ln(R/r_{dc}) + W_{dc}$, where $R$ is a characteristic dimension of the considered system, $L$ is the disclination length, $r_{dc} \sim 10$ nm and $W_{dc}$ are the radius and the energy of the disclination core, respectively.[19] For $R_{np}$ comparable to the defect core size, the energy reduction is mostly due to the energy of the displaced isotropic defect core, but the contribution of the elastic part of the defect energy increases further with increasing $R_{np}$. Using the calculated reduction energy $\Delta W_R$ for different nanoparticles and defects (Table 2), one can roughly estimate the order of magnitude of the corresponding trap escape forces as $F_{te} \sim \Delta W_R/R_{eqs}$ (Table 2). Consistent with the range of obtained values, optical tweezers at moderate laser powers of about 50 mW (corresponding optical trapping forces of up to ~20 pN[27]) at the sample plane do not exert optical trapping forces that would be strong enough to remove nanoparticles from the topological defect traps once they are entrapped. The stiffness of these topological defect traps is $\kappa_{tdt} \geq k_B T/\Delta r^2$, where $\Delta r \sim 10$ nm is the maximum experimentally observed displacement of the nanoparticle from the center of the defect trap, precise measurement of which is limited by the resolution of videomicroscopy. The estimated stiffness $\kappa_{tdt}$ for different studied nanoparticles is 20−100 pN/nm, which is significantly larger than the stiffness of optical tweezers trapping gold nanoparticles at relatively high laser powers[15−17] and plasmonic optical traps.[45]

The mobility of nanoparticle-microsphere dimers in the LC can be used to assemble desired photonic structures. Figure 6 shows examples of such assemblies levitating in the LC bulk because of the repulsive elastic interactions between the colloidal dimers and confining substrates. Once nanoparticles are trapped by defects, the dimers of plasmonic and dielectric particles can assemble into patterns dictated by elastic interactions in the LC. For example, Figure 6a−c shows a pair of MGs trapped into the hedgehogs due to two microspheres. Figure 6d−g shows a chain of MG-microsphere pairs along $\mathbf{n}_0$; the 3 μm spacing between MGs is defined by the diameter of microspheres. Nanoparticles can be also assembled into two-dimensional arrays, such as the one composed of SF nanoparticles trapped into boojums (Figure 6h,i). Interestingly, the alignment of SFs often changes as multiple microparticles self-assemble into more complex structures, which may be promoted by sharing of individual anisotropic nanoparticles by two boojums induced by the two interacting microspheres (Figure 6i). The distance between nanoparticles and their alignment in such structures can be controlled by using microspheres with different size and surface chemistry, depending on the need of specific photonic applications. Although the surface boojums localize nanoparticles in direct contact with the surface of the microspheres, the bulk point and line defects can localize nanoparticles at a well-controlled distance and orientation with respect to the surface of the microsphere. The distance from the center of microsphere to the hyperbolic point defect is known to be ~$1.2a$,[38,39] where $a$ is the radius of the microsphere, while the radius of the Saturn ring defect is typically ~$1.1a$.[41] By controlling the size of the colloidal microspheres in the submicrometer and micrometer ranges as well as controlling the size and shape of entrapped nanoparticles, one can use the demonstrated scaffolding of nanoparticles to form two-particle nanoantennas with the distance between the surfaces of bigger and smaller colloids controlled within 10−100 nm. One can also envision the assembly of ring-shaped structures of multiple plasmonic nanoparticles entrapped and oriented in the Saturn ring defect line encircling a bigger metallic or dielectric colloidal sphere. In addition to microspheres, one can utilize bigger particles of other shapes, like platelets and rings that are known to induce different types of defects.[44] Importantly, all of these fluid-borne colloidal microparticles with scaffolded nanoparticles can be further used for self-assembly or light-guided hierarchical superstructures with a host of potential applications, such as fabrication of novel types of tunable metamaterials.

In conclusion, we have demonstrated oriented trapping of plasmonic gold nanoparticles by topological singularities in nematic LCs. The defect traps induced by colloidal microspheres allow for highly precise, oriented spatial localization, strong maximum trapping potentials of 100−5000 $k_\mathrm{B}T$, and large trap stiffnesses that depend on the shape and size of trapped nanocolloids and are of great interest for fundamental studies in nanophotonics and plasmonics. The demonstrated elastic scaffolding of smaller nanoparticles by elastic distortions and defects due to bigger microparticles yields an interesting colloidal system with the anisotropic nanoparticles at well-defined orientation with respect to the surface of microparticles and the far-field LC director. The ensuing colloidal dimers of plasmonic metal nanoparticles and dielectric microparticles can be assembled into a number of one- and two-dimensional arrays of desired configuration, as needed for applications in nanoscale photonics and plasmonics. The use of metal microparticles instead of dielectric ones may enable applications in nanoantennas as well as in nanoscale energy conversion systems. Similar approaches may be used for nanoscale trapping of other types of nanoparticles, like semiconductor nanocrystals, which are important for imaging and energy conversion. Since self-assembled structures of plasmonic nanoparticles can exhibit a broad range of magnetic and electric resonances,[7] plasmonic nanoparticle self-assemblies in a fluid LC medium with facile response to electric fields may offer a means for low voltage control of their optical response and enable self-assembly based fabrication of tunable bulk metamaterials.

**Acknowledgments**

This work was supported by the Division of Chemical Sciences, Geosciences, and Biosciences, Office of Basic Energy Sciences of the U.S. Department of Energy under Contract No. DEAC36-08GO28308 with the National Renewable Energy Laboratory (J.vdL and J.S.E.), by the International Institute for Complex Adaptive Matter (B.S.), and by NSF Grants DMR-0847782, DMR-0820579, and DMR-0844115 (I.I.S., B.S., P.J.A., and T.L.). E.R.Z. acknowledges financial support by NSF (DMR-0547399, DMR-1105878). We thank Christian Schoen and Shelley Coldiron from Nanopartz, Inc. for providing gold nanoparticles along with the corresponding TEM images (Figure 1a,c,d). We acknowledge discussions with Nick Abbott, Noel Clark, and Victor Pergamenshchik.

## Figures

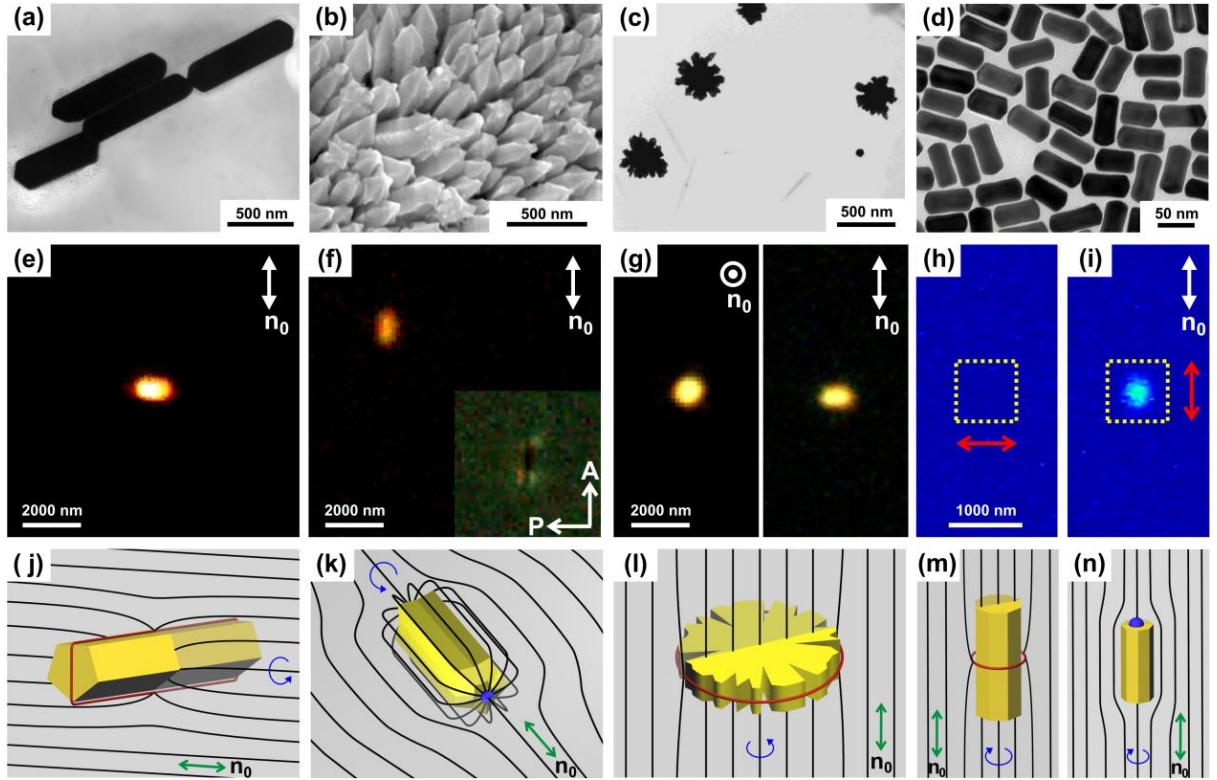

**Figure 1.** Gold nanoparticles in a nematic LC. (a,c,d) TEM images of MG, NB and GNR nanoparticles, respectively. (b) SEM image of SF nanoparticles. (e−g) Dark-field microscopy images of MGs, SFs, and NBs in 5CB, respectively. Inset in (f) shows the texture of director distortions around SF between crossed polarizer "P" and analyzer "A". (h,i) TPL images of an sGNR nanoparticle (in the area outlined by yellow dashed line) taken with the polarization of excitation beam ($\lambda_{exc}$ = 850 nm) perpendicular (h) and parallel (i) to $\mathbf{n}_0$. (j-n) Schematic diagrams of $\mathbf{n}(\mathbf{r})$ (black lines) around MG, SF, NB, GNR, and sGNR nanoparticles, respectively. Red lines in (j,l,m) and blue semispheres in (k,n) show a $k = -1/2$ defect line and an $m = -1$ boojum point singularities, respectively. Green and white double arrows show the in-plane direction of the far-field director $\mathbf{n}_0$; white circle with the point in the middle show $\mathbf{n}_0$ normal to the field of view. Red double arrows in (h,i) show the polarization direction of the excitation beam. Blue arrows in (j−n) show the allowed rotations of aligned nanoparticles.

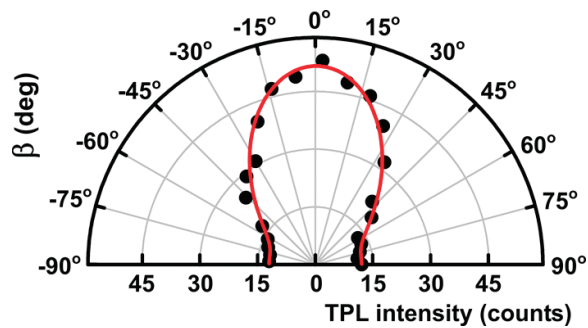

**Figure 2.** TPL intensity from a sGNR nanorod in a nematic LC vs the angle $\beta$ between its long axis and the polarization of excitation beam, revealing its alignment along $\mathbf{n}_0$. The red line shows the best fit of experimental data (black filled circles) with the expected angular dependence of TPL $\sim \cos^4\beta$.

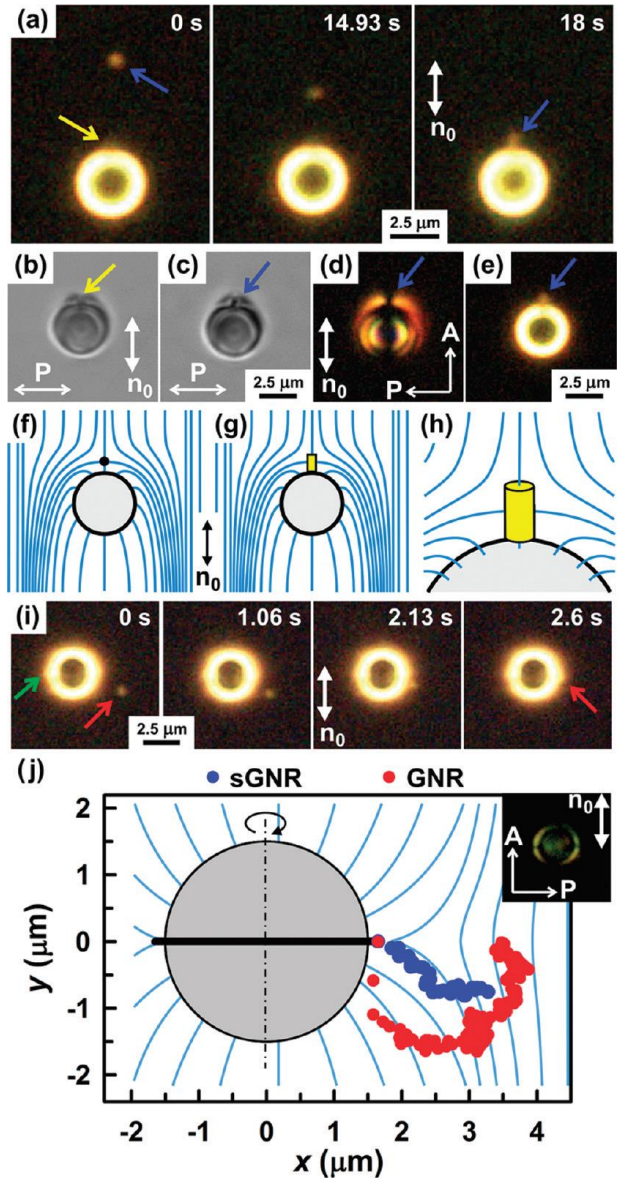

**Figure 3.** Attraction of GNRs and sGNRs to topological singularities in a nematic LC. (a) A sequence of dark-field images showing an sGNR nanorod (marked by a blue arrow) moving into an $N = -1$ point defect (marked by a yellow arrow) near the microsphere. (b) A bright-field image of the microsphere with the point defect. (c−e) Bright-field, polarizing, and dark-field images, respectively, of silica microsphere with the point defect displaced by an sGNR. (f) A schematic diagram of the dipolar $\mathbf{n}(\mathbf{r})$ around the microsphere; black filled circle shows the point defect. (g) A schematic diagram of n(r) with the nanorod (yellow) displacing the defect. (h) An enlarged schematic view of the nanorod (yellow cylinder) in the defect. (i) Dark-field images showing attraction of a GNR (marked by a red arrow) to a Saturn ring defect (marked by a green arrow). (j) Trajectories of GNR (red filled circles) and sGNR (blue filled circles) nanoparticles attracting to the "Saturn ring" (black line). $\mathbf{n}_0$ is shown by white double arrows and $\mathbf{n}(\mathbf{r})$ in (f−h,j) is shown by blue lines. Inset in (j) shows the polarizing microscope texture of the microsphere with the Saturn ring defect.

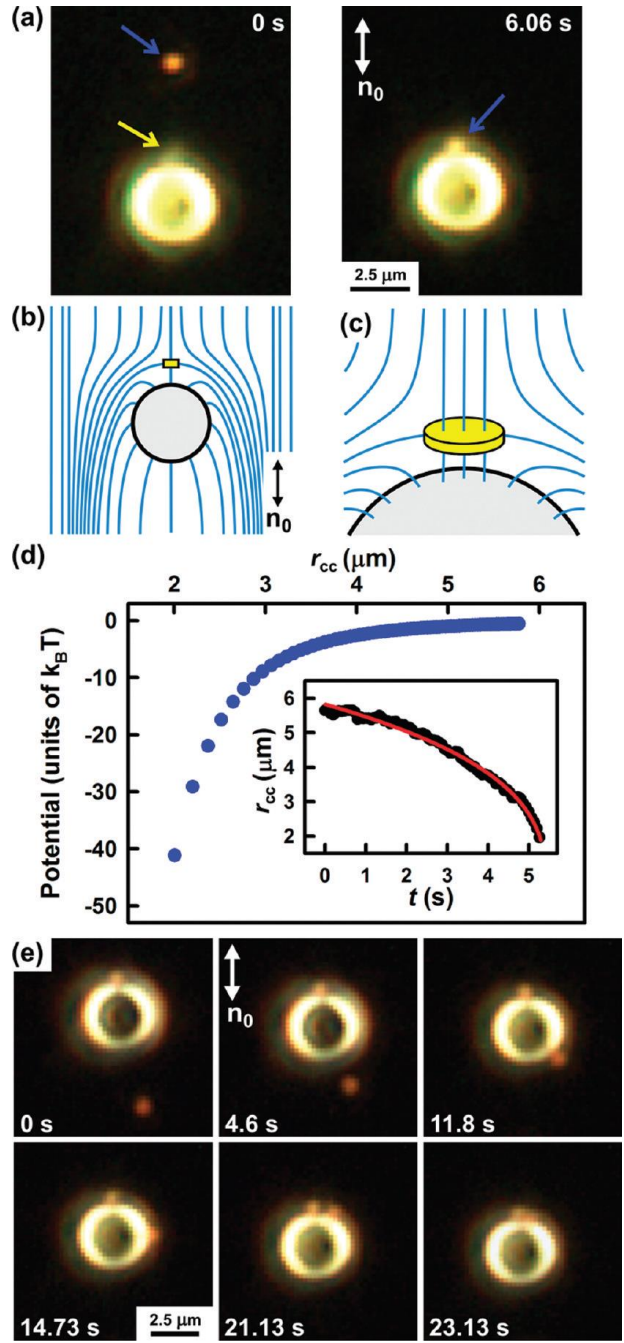

**Figure 4.** Attraction of a NB nanoplatelet to a microsphere-induced point defect in a nematic LC. (a) Dark-field images show the NB (marked by a blue arrow) moving toward the point defect (marked by a yellow arrow). (b) A schematic diagram of dipolar $\mathbf{n}(\mathbf{r})$ with NB nanoparticle (yellow) displacing the defect. (c) An enlarged schematic view of the NB platelet in the trap. (d) The elastic interaction potential as a function of a center-to-center separation between particles $r_{cc}$, derived from the inset plot of $r_{cc}$ vs time $t$. (e) Dark-field video frames showing motion of NB to a point defect from the initial position "behind the microsphere," with its sliding around the microsphere surface toward the defect.

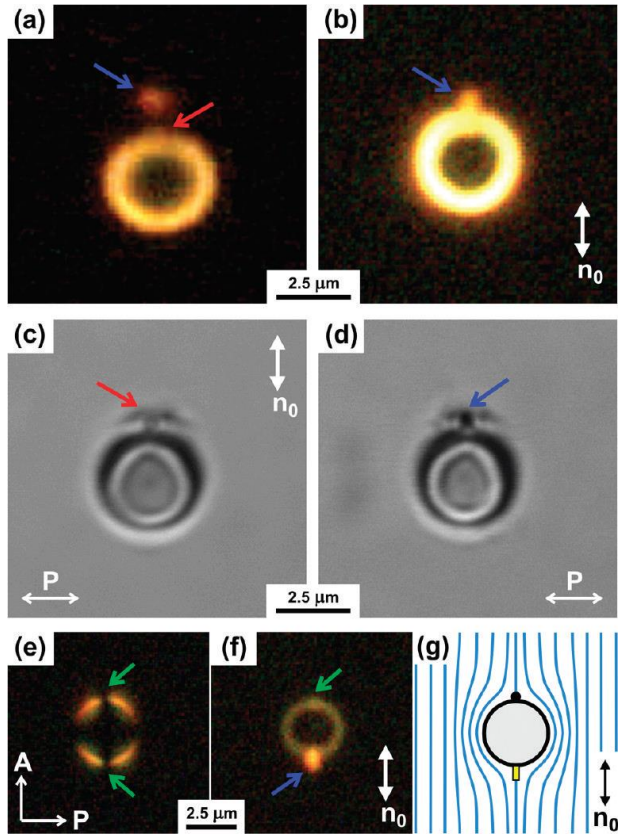

**Figure 5.** Entrapment of MGs and SFs by point singularities assisted by optical tweezers. (a) Dark-field image showing MG (marked by a blue arrow) in the initial local-minimum-energy location away from a bulk point defect (marked by a red arrow). (b) A dark-field image of an MG nanoparticle stably trapped in a point defect region after the use of optical tweezers to overcome the elastic energy barrier. (c) A bright-field image of the microsphere with dipolar $\mathbf{n}(\mathbf{r})$ and a point defect. (d) A bright-field image of an MG nanorod (pointed by a blue arrow) displacing the bulk point defect. (e) A polarizing microscopy image showing the quadrupolar $\mathbf{n}(\mathbf{r})$ around melamine resin microsphere. (f) A dark-field image of SF nanorod (marked by the blue arrow) entrapped by a surface point defect (the boojum is marked by a green arrow) with the assistance of optical tweezers. (g) A schematic of $\mathbf{n}(\mathbf{r})$ (blue lines) around a microsphere with tangential anchoring and SF nanoparticle (yellow) corresponding to (e) and (f); the black filled semicircle in (g) shows the boojum.

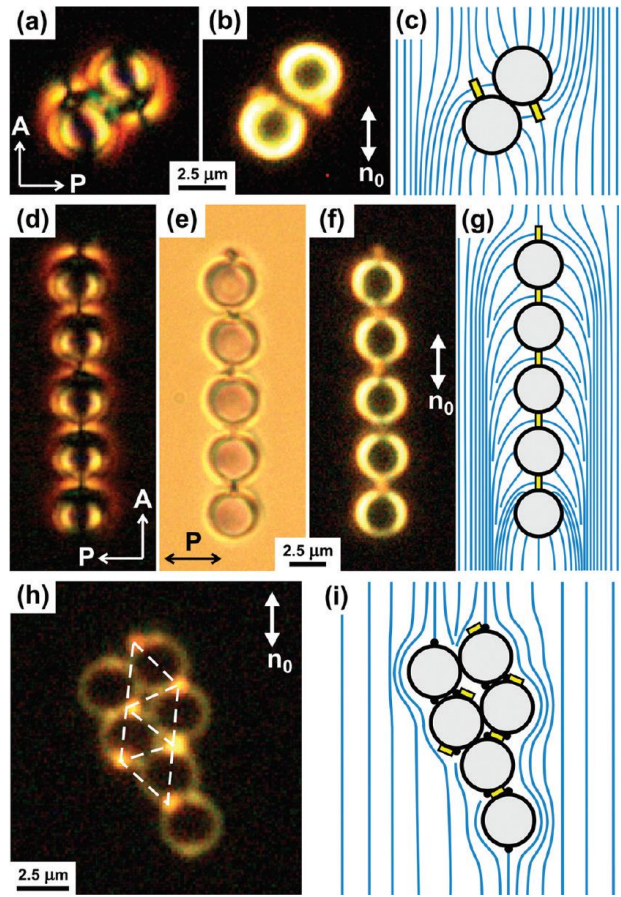

**Figure 6.** Colloidal superstructures of gold nanoparticles and dielectric microspheres. (a−c) Polarizing and dark-field microscopy images and a schematic diagram showing microsphere-MG pairs assembled in antiparallel direction, respectively. (d−g) Polarizing, bright-field, and dark-field microscopy images and a schematic diagram showing chains of microsphere-MG dimers aligned along $\mathbf{n}_0$, respectively. (h,i) A darkfield image and a schematic diagram of an array (accentuated by a dashed line) formed by SF nanorods trapped in distorted regions next to melamine resin microspheres.

**Table 1.** Properties of gold nanoparticles in a nematic liquid crystal

| nanoparticles | capping | axial diameter $D_{np}$ (nm) | length $L_{np}$ (nm) | elastic energy of distortions in an "equivalent" sphere $\Delta W$ ($k_B T$) | attraction energy near contact with point/line defects $W_0$ ($k_B T$) |
|---|---|---|---|---|---|
| Microgold (MG) | NSol(akyl acrylate) | 150 | 800 | 235 | - |
| Starfruit (SF) | NSol(akyl acrylate) | 100 | 500 | 153 | - |
| Nanoburst (NB) | NSol(akyl acrylate) | ~500 | 100 | 262 | ~40/- |
| nanorods (GNR) | NSol(akyl acrylate) | 25 | 60 | 30 | (3-5)/20 |
| nanorods (sGNR) | polystyrene | 10 | 45 | 15 | (3-5)/20 |

**Table 2.** Properties of topological defect traps for different gold nanoparticles

| nanoparticles | reduction energy $\Delta W_R$ ($k_B T$) | | trap escape force $F_{te}$ (pN) | |
|---|---|---|---|---|
| | point defect | line defect | point defect | line defect |
| MG | 1854 | 3817 | 51 | 105 |
| SF | 1170 | 2335 | 49 | 99 |
| NB | 2078 | 4318 | 51 | 107 |
| GNR | 132 | 334 | 29 | 73 |
| sGNR | 59 | 136 | 27 | 62 |

# Supporting Information

### 1. Two-photon luminescence of gold nanorods sGNR and NB nanoparticles

We determined the orientation of small nanorods sGNR and GNR using a polarization dependent two-photon luminescence (TPL) of gold nanorods.[S1,S2] Here, as an example, we present the study of TPL from sGNRs (dimensions 45 × 10 nm). Nanorods were spin-coated from an organic solvent solution on a clean glass substrate. Two glass substrates were assembled into the cell with a gap set by nanorods.

A tunable (680-1080 nm) femtosecond Ti:sapphire oscillator (140 fs, 80 MHz, Chameleon Ultra-II, Coherent) was used to generate TPL from sGNRs. The excitation of gold sGNRs was performed at wavelengths corresponding to their longitudinal surface plasmon resonance peak ($\lambda_{exc}$ = 850 nm) and the TPL signal was detected within the range 400-700 nm by photomultiplier tube (H5784-20, Hamamatsu). The in-plane position of the excitation beam was controlled by a laser scanning unit (FV300, Olympus) coupled to the inverted Olympus microscope IX81 capable of dark-field imaging as well. The polarization of excitation beam was changed using the half-wave retardation plate mounted before an 100× (NA = 1.4) oil objective, which both focused the excitation light into the sample and collected light emitted from nanorods. To measure the TPL spectrum we used a fiber optic spectrometer USB2000 (Ocean Optics, Inc.) and the average excitation power $P_{IN}$ was in the range 10-100 mW at the laser output.

We measured the dependence of TPL intensity on the average excitation power (Fig. S1). The TPL intensity increased as the excitation power increased (Fig. S1a). The slope of the linear fit in Fig. S1b was 1.95, which is indicative of two-photon excitation process.[S1,S2]

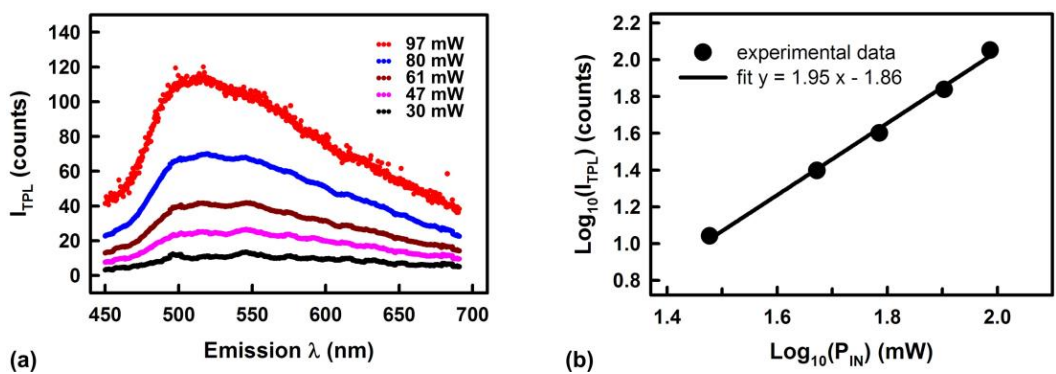

**Figure S1.** Dependence of TPL intensity on excitation power: (a) TPL spectra at different excitation powers; (b) Log-Log plot of TPL intensity vs. excitation power. The slope of the fit is 1.95, which indicates a two-photon excitation process.

For TPL imaging, the power of excitation beam before focusing by objective was just ~100-200 μW. Scanning the area of the cell with the excitation beam we obtained TPL images of single nanorods (Fig. S2). The TPL from nanorods is excitation polarization dependent.[S1,S2] Figure S2 shows TPL images taken for two orthogonal polarizations of excitation beam. The brightest spots correspond to nanorods aligned parallel to the polarization of excitation beam and, from a superimposed image (Fig. S2c), one can distinguish the orthogonally oriented nanorods.

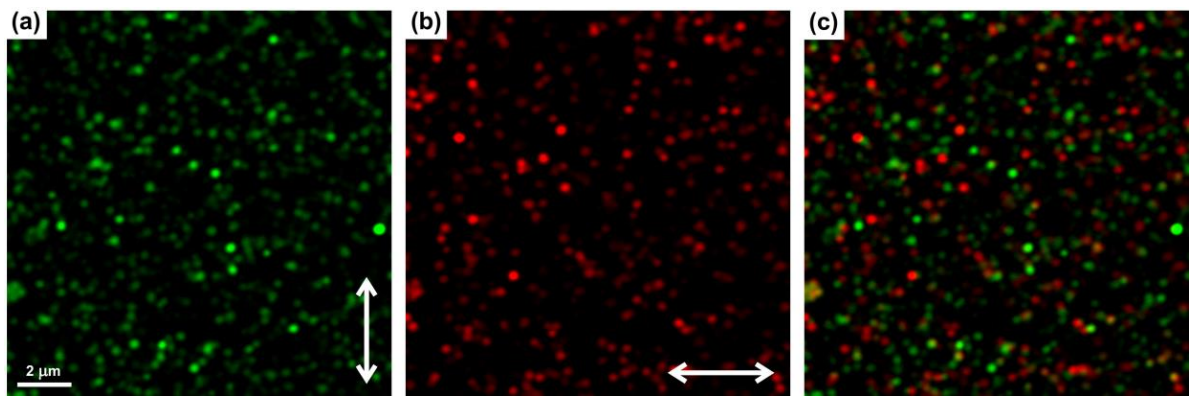

**Figure S2.** TPL images of nanorods sGNR spin-coated on glass cover slip taken at two orthogonal polarizations of excitation beam: (a) TPL image at vertical polarization; (b) TPL image at horizontal polarization; (c) superimposed image of (a) and (b). White double arrow shows the polarization of excitation beam. Excitation wavelength was 850 nm.

It is relatively difficult to determine the orientation of NB platelets at the surface of microsphere using the polarized dark-field microscopy as the scattering from microsphere usually overshadows the scattering from NB. Therefore, we also use TPL imaging with excitation at 850 nm to verify the orientation of NB nanoparticles nearby the microsphere surface (Fig. S3). The plasmonic response of NB is rather complicated due to the complex irregular shape of their edges. Nevertheless, the intensity of emission from NB is strongly dependent on the polarization of the excitation beam (Fig. S3). The emission intensity is maximum for the laser excitation with linear polarization perpendicular to the director (parallel to the flat surface of NB) (Fig. S3a) and minimum when the polarization of excitation is parallel to the director (perpendicular to the flat surface of NB) (Fig. S3b). This, along with the consideration of the known surface boundary conditions for the LC director, allowed us to conclude that the flat surfaces of NB nanoparticles orientate parallel to the surface of colloidal microsphere.

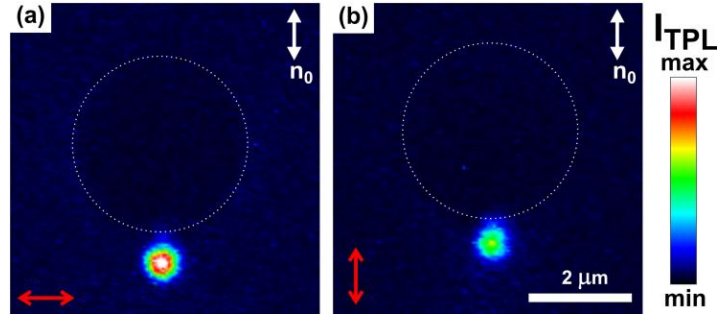

**Figure S3.** TPL images of the NB nanoparticle elastically entrapped in the hedgehog defect region created by a microsphere with vertical surface boundary conditions. (a,b) The images are obtained for two orthogonal polarizations of the excitation laser light: (a) TPL image at horizontal polarization; (b) TPL image at vertical polarization. Red double arrow shows the polarization of excitation laser beam. Excitation wavelength was 850 nm. White dashed circle delineates the boundary of the microsphere. The area inside the dashed circle is darker than the outside area with liquid crystal due to the weak three-photon self-fluorescence from the used LC material.[S3] The color bar shows the color-coded scale of intensity $I_{TPL}$ of the detected TPL signal.

## 2. Interaction between elastic quadrupoles formed by colloidal microspheres and GNR and sGNR nanorods

Colloidal particles introduced into the uniform nematic LC cell cause distortions of director field $\mathbf{n(r)}$[S4,S5] that can have dipolar or quadrupolar symmetry. There are two types of elastic quadrupoles observed in our experiments. Elastic quadrupoles with a disclination loop called "Saturn ring" are formed by microspheres and GNR nanorods (Fig. S4b). The elastic quadrupole with two surface point defects called "boojums" (Fig. S4a) is formed by sGNRs. The colloids with "Saturn ring" and "boojums" have opposite distributions of signs of defects in their quadrupolar structures.[S6] The directionality of elastic forces between two elastic quadrupoles depends on whether they have similar or opposite distributions of the defect signs. Figure S4a schematically shows that colloids with a "Saturn ring" and "boojums" have elastic quadrupoles of the opposite defect sign distributions and opposite curvatures of elastic distortions; therefore, they exhibit the strongest attraction when separated along the directions normal and parallel to the far field director $\mathbf{n_0}$; particles repulsion takes place at intermediate angles.[S6] However, microsphere and GNRs have the same curvatures of elastic distortions and distributions of defects (Fig. S4b) and, thus, repel for center-to-center separations along and normal to $\mathbf{n_0}$ but attract at intermediate angles.[S4,S5] This behavior can be qualitatively understood using simple considerations of minimization of elastic energy due to matching or mismatching of elastic distortions as the nanoparticles approach the colloidal microsphere from different directions (Figure S4).

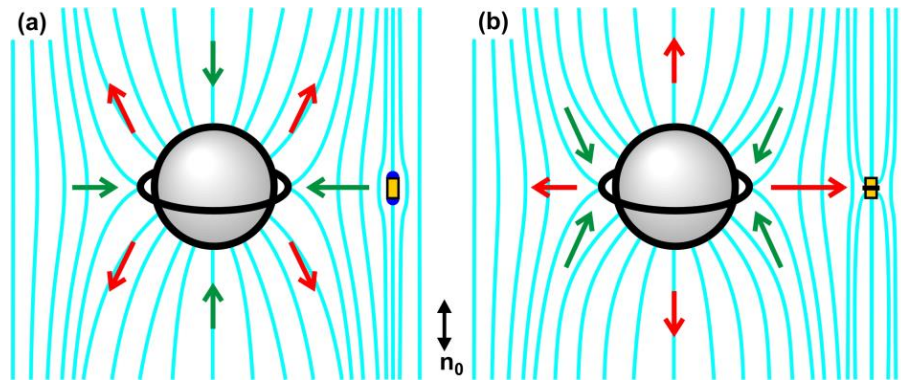

**Figure S4.** Schematic diagrams showing directionality of interactions between elastic quadrupoles induced by a microsphere and (a) sGNR and (b) GNR nanorods (yellow rectangles). Cyan lines and a double black arrow show the director field **n(r)** and far field director **n$_0$**, respectively. Green and red arrows show the directions of attraction and repulsion of the nanoparticles, respectively. Two blue semispheres in (a) show two surface point defects boojums at the ends of sGNR. Black line around a sphere and a black line across a yellow rectangle in (b) show disclination loops called "Saturn rings".